\documentclass[12pt]{iopart}
\usepackage{graphicx}
\usepackage{color}

\begin{document}

\title{Spatially addressable readout and erasure of an image in a gradient echo memory}{}

\author{Jeremy B. Clark$^*$, Quentin Glorieux and Paul D. Lett}

\address{Quantum Measurement Division, National Institute of Standards and Technology\\
and Joint Quantum Institute, NIST and University of Maryland,\\
100 Bureau Dr., Gaithersburg, MD 20899-8424\\
$^*$Contact address: jeremyc@nist.gov}
\pacs{42.50.Md,42.79.Vb,71.36.+c}

\begin{abstract}
We show that portions of an image written into a gradient echo memory can be individually retrieved or erased on demand, an important step towards processing a spatially multiplexed quantum signal.
Targeted retrieval is achieved by locally addressing the transverse plane of the storage medium, a warm $^{85}$Rb vapor, with a far-detuned control beam.
Spatially addressable erasure is similarly implemented by imaging a bright beam tuned near the $^{85}$Rb D$_1$ line in order to scatter photons and induce decoherence.
Under our experimental conditions atomic diffusion is shown to impose an upper bound on the effective spatial capacity of the memory.
The decoherence induced by the optical eraser is characterized and modeled as the response of a two level atom in the presence of a strong driving field.
\end{abstract}

\maketitle

\section{Introduction}
Quantum communication protocols are based on sharing entanglement between remote locations \cite{Braunstein:2005wr,Kimble:2008uv}.
Since photons are often used in the transportation of quantum states, optical loss typically limits the maximum achievable distance over which entanglement can be shared.
Several proposals have addressed this problem by dividing the quantum channel into shorter segments that are separately purified and connected by entanglement swapping \cite{Sangouard:2011bp}.
These protocols rely on the availability of a quantum memory allowing for the storage and retrieval of quantum information.

A variety of storage methods have been developed in atomic systems, many of which exploit collective excitations in atomic vapors \cite{Lvovsky:2009vg}.
Using a vapor of $^{87}$Rb, the Gradient Echo Memory (GEM) technique has recently been shown to yield retrieval efficiencies of up to 87\% and recall fidelities exceeding the no-cloning limit \cite{Hosseini:2011iv}.
GEM is based on the coherent transfer of a weak optical pulse via Raman coupling to a spin wave excitation in the hyperfine ground state of the storage medium \cite{Hosseini:2012go}.
The Raman transition is broadened due to the Zeeman effect by applying a magnetic field gradient along the propagation axis.
The inhomogeneity of the magnetic field causes the excited longitudinal spin wave to reversibly dephase.
Inverting the slope of the gradient after a brief period causes the spin wave to begin rephasing.
Applying an intense Raman-coupled control beam coherently transfers the rephased spin wave back into an optical echo which emerges in the forward direction.

Extending the storage time and recall fidelity of this system is a challenging problem since the coherence time of the ground state ultimately limits the performance of the memory \cite{Hetet:2008wi}.
It has been shown, however, that storage time requirements for quantum repeaters can be reduced by several orders of magnitude if a multiplexed memory is used \cite{Collins:2007cs, Lan:2009ts}.
Since GEM has been shown to allow spatiotemporal multiplexing \cite{Glorieux:2012uo, Higginbottom:2012gl}, it is a good candidate to accommodate spatially multiplexed signals sent between nodes of a quantum repeater.
Narrowband single-photon sources compatible with this memory have already been demonstrated \cite{Lvovsky:2012}, and spatially multiplexed signals consisting of only a few photons have been stored using similar techniques in a cold atomic gas \cite{Guo:2013}.
In addition to storing single photons, spatial multiplexing would aid in the distribution of continuous-variable entanglement.
For example, entangled Gaussian states shared between nodes could be distilled by way of heralded non-Gaussian operations \cite{Eisert:2002} such as photon subtraction \cite{Furusawa:2010} before being selectively connected by entanglement swapping \cite{Zhang:2011}.

In this paper, we demonstrate that it is possible to selectively retrieve, at different times, subregions of an image stored in the memory.
We store an image of the NIST logo and partially retrieve 3 different subregions of the image after 2, 2.5, and 3 $\mu$s of storage time.
This suggests that the selective retrieval of multi-spatial-mode quantum states should in principle be possible using a multiple--rephasing protocol similar to \cite{Hosseini:2009fd}.
To enhance the flexibility of this partial retrieval protocol, we also introduce a technique that allows for selective erasure of subregions of the image during storage.
By applying an intense beam referred to as an ``optical eraser" near the D$_1$ resonance (795 nm), we can rapidly induce decoherence in the spin wave through spontaneous emission.
If the optical eraser is confined in the transverse plane of the memory, an isolated subregion of the image can be targeted for deletion.
For a multi-spatial-mode quantum state, it should be possible to selectively erase a limited number of quantum channels during storage without degrading the quantum states of neighboring modes.

\subsection{Experimental set-up}
\begin{figure}[htbp]
\centering\includegraphics[width=.9\columnwidth]{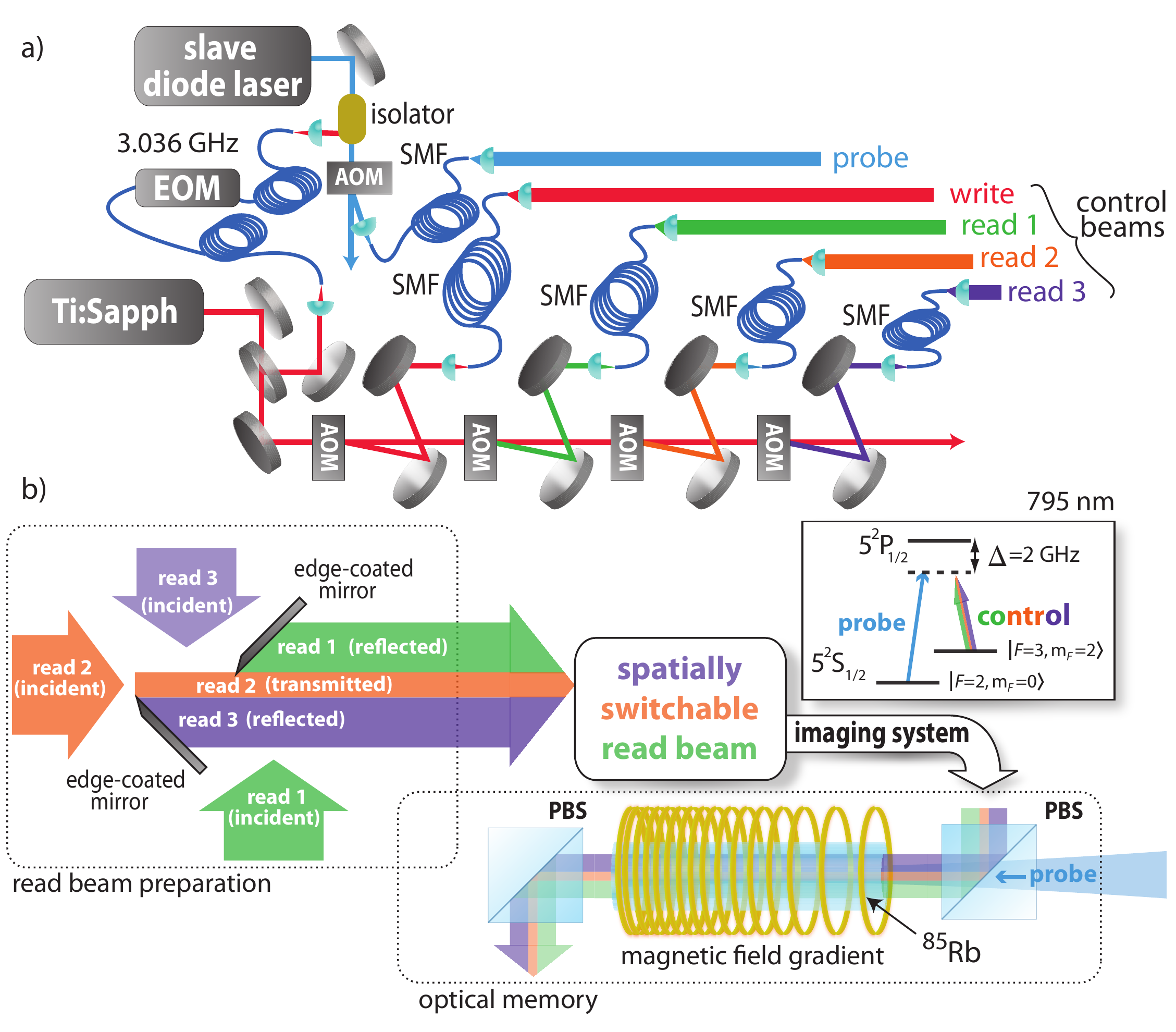}
\caption{\label{fig:setup} Experimental setup for multiple--readout GEM experiment.
a) Generation of the probe, write, and individual read beams.
A titanium-sapphire (Ti:Sapph) laser serves as a master laser near the control beam frequency.
An electro-optic phase modulator (EOM) driven at 3.036 GHz is used to generate optical sidebands in the master laser light, and a slave diode laser is injection locked at the first blue sideband to prepare the probe beam.
The probe and control beams are amplitude-modulated using independent acousto-optic modulators (AOMs) and spatially filtered with single mode fibers (SMFs).
b) The three read beams are combined using edge-coated mirrors and imaged into the memory cell.  The probe and control beams are combined using a polarizing beam splitter (PBS).  Inset: Raman coupling of the ground state near the $^{85}$Rb D$_1$ line.}
\end{figure}

Our apparatus allows three different sub-regions of the atomic excitation to be read out independently (Fig.~\ref{fig:setup}).
From a master Ti:sapphire laser, a fraction of the light is sent through a fiber-coupled electro-optic phase modulator driven at 3036 MHz corresponding to the hyperfine splitting of the ground state in $^{85}$Rb.
The first blue optical sideband of the modulated light is selected by injection locking a current-tuned diode laser.
The diode light is then pulsed using an 80~MHz acousto-optic modulator (AOM), and the first diffracted order of the AOM is spatially filtered using a single mode fiber.
Finally, the mode shape of the light emerging from the fiber is expanded with a telescope and passed through a 2$\times$8~mm binary intensity mask of the NIST logo.
The mask is imaged into the atomic memory with a magnification of 1.25.
We refer to this blue-detuned image as the probe.

The remaining master laser power is sent through a chain of four additional AOMs all driven at 80 MHz, and the first diffracted order from each AOM is coupled into a single mode fiber (Fig.~\ref{fig:setup}-a).
Independently driving these AOMs allows us to implement an arbitrary pulse sequence among the beams while maintaining control of their optical frequencies.
We use one of these control beams, referred to as the ``write" beam, to mediate the Raman coupling of the probe pulse to the atomic ground state coherence.
This 50 mW write beam is spatially filtered using a single mode fiber, collimated to a $\frac{1}{e^2}$ diameter of 2 cm, and combined with the probe beam on a polarizing beamsplitter similar to \cite{Glorieux:2012uo}.

The three remaining ``read" control beams are included in the setup to address different subregions of the transverse plane of the vapor cell at different times.
These beams are prepared at the same optical frequency to ensure that all retrieved pulses emerge at the same frequency.
This approach simplifies any desired homodyne detection in future experiments by requiring only a single local oscillator frequency to detect the field quadratures of each retrieved subregion.
To achieve time-dependent spatial addressability, the optical paths of the beams are combined using an assembly of edge-coated mirrors as illustrated in Fig.~\ref{fig:setup}-b.
The combination plane is imaged into the cell in order to obtain two sharply defined boundaries between the three zones of the image that are to be independently retrieved at different times.

In all experiments, we Zeeman-broaden the Raman line to a width of $\approx$~1~MHz.
All probe pulses were temporally shaped as Gaussians with a $\frac{1}{e^2}$ width of 2~$\mu$s.
The write beam is switched off during 2~$\mu$s of storage time, and portions of the image are sequentially retrieved using three read beams each pulsed for 500~ns during the rephasing time of the spin wave (Fig.~\ref{fig:fig2}a).
The atomic memory itself consisted of a 20~cm long $^{85}$Rb cell with 1.33~kPa (10~Torr) of Kr buffer gas heated to approximately 80$^{\circ}$C.
Similar to \cite{Vudyasetu:2008wv,Shuker:2008bn}, we choose a relatively high partial pressure of the Kr buffer gas in order to mitigate the blurring effect observed on stored images due to atomic diffusion \cite{Glorieux:2012uo,Higginbottom:2012gl}.
Using a technique identical to \cite{Glorieux:2012uo}, we estimate a diffusion coefficient for these experiments to be 35 cm$^2$s$^{-1}$.
The echoes are captured using a gated intensified CCD camera in order to study the time dynamics of the readout.

\section{Multi-spatial-mode readout}
\begin{figure}[htbp]
\centering\includegraphics[width=\columnwidth]{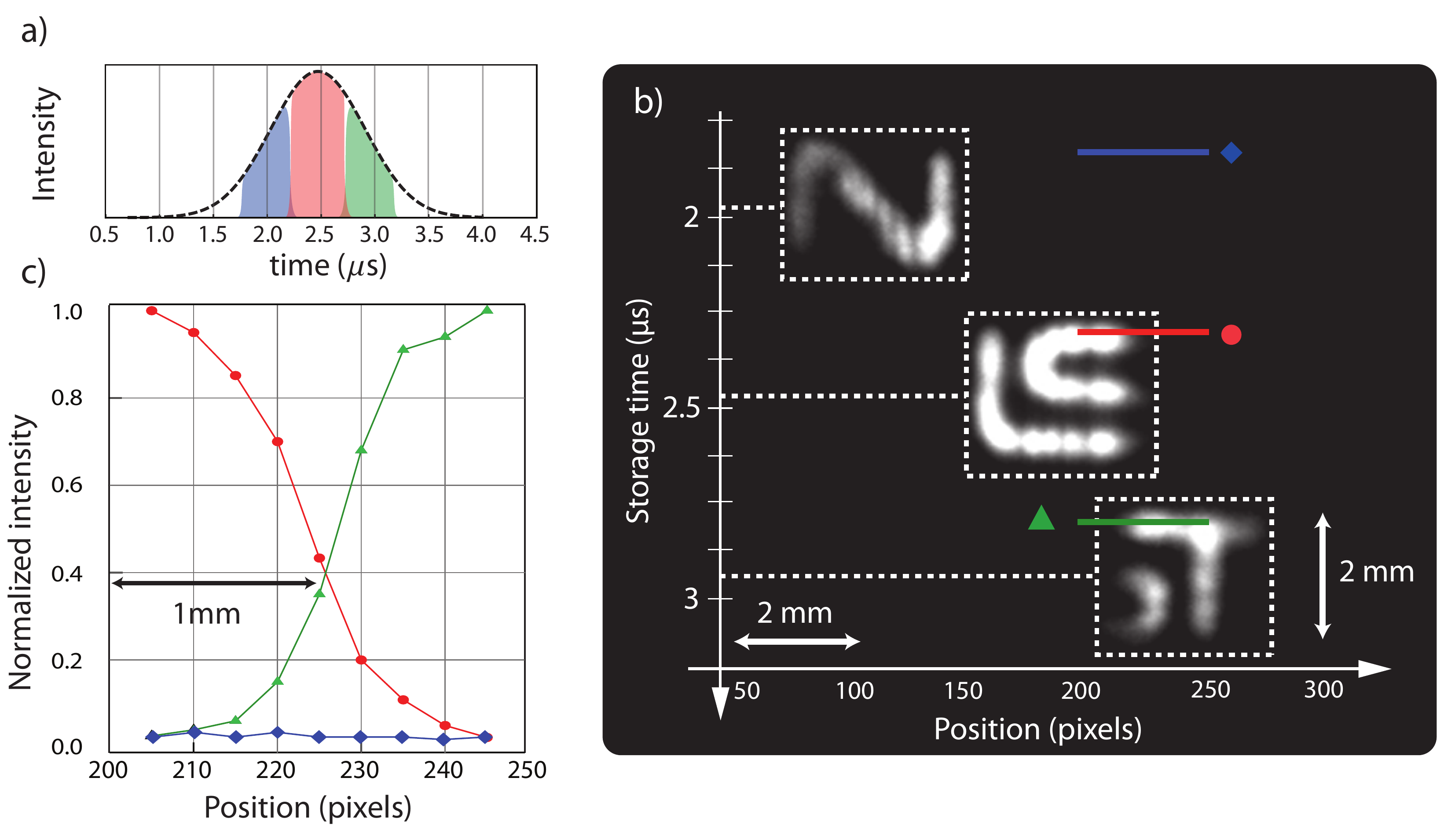}
\caption{\label{fig:fig2} a) Retrieval sequence as a function of storage time.  Regions shaded blue, red, and green respectively denote the first, second, and third recalled subregions of the stored image.
The black dotted line indicates the typical temporal shape of a single retrieved echo.
b) Retrieved portions of the NIST logo as function of storage time and horizontal position.
The NIST logo was stored once as a single input using a single pulse of the write beam.
The image was then retrieved piecewise at various times (2, 2.5 and 3~$\mu$s) using three different read beams separately activated during the approximately 3~$\mu$s of rephasing time of the spin wave.
Part a) of the figure indicates the fraction (in time) of the stored echo that is read out in the different spatial regions.
c) Profile plot of the normalized intensity for a selected subset of pixels in order to show the boundary between two read-out regions: blue diamonds denote the retrieval at 2~$\mu$s, red circles at 2.5~$\mu$s, and green triangles at 3~$\mu$s.}
\end{figure}

The longitudinal orientation of the spin wave allows any transverse intensity modulation of the stored pulse to be recovered in the echo \cite{Glorieux:2012uo}.
Similarly, we are free to modulate the transverse intensity profile of the read beam in order to selectively read out any desired portion of a stored image.
Three subregions of the NIST logo retrieved by the spatially addressable read beams are portrayed in Fig.~\ref{fig:fig2}b.
These subregions were retrieved piecewise during a single $\approx$~2~$\mu$s rephasing process and not from rephasing the spin wave three times \cite{Hosseini:2009fd} or performing three separate echo experiments.

It is apparent from Fig.~\ref{fig:fig2}a that the second retrieved subregion of the image is brighter than the first and last retrievals.
This effect is due to nonuniform illumination of the mask with the probe beam and to the temporal shape of the retrieved pulse.
Although the probe beam was expanded and collimated before passing through the intensity mask, its Gaussian profile contributed to a 35\% reduction in intensity near the edges of the mask.
Since the temporal shape of the echo is also Gaussian with a maximum intensity corresponding to a retrieval time of 2.5~$\mu$s, the first and last retrievals exhibited up to an additional 20\% reduction in intensity.

To characterize the addressability of the memory associated with this technique, we include plots of the normalized intensity profile for a single row of pixels at three different readout times in Fig.~\ref{fig:fig2}b. 
The profile plots illustrate the extent to which a continuous intensity profile (the connection between the letter ``S" and the letter ``T") written into the memory can be sharply separated between two readouts.
The normalized intensity profiles recalled in two neighboring readouts show a decline from 0.9 to 0.1 over a length of $\approx$~900~$\mu$m.
This length scale is consistent with that of the intensity profiles of the read beams imaged into the cell, confirming that the readout resolution is limited by the sharpness of the boundaries between the read beams.

Our strategy of applying several localized read beams during a single rephasing process introduces an effective loss for each individually--recalled subregion of the stored image.
In the context of storing and retrieving a spatially multiplexed quantum signal such loss can be avoided by repeatedly inverting the magnetic field gradient as shown in \cite{Hosseini:2009fd} and applying the localized read beams during the entirety of each rephasing time.
This approach would extend the total recall time for the experiment, rendering the blurring effects due to the diffusive motion of the atoms to be more significant during each successive targeted retrieval \cite{Glorieux:2012uo, Firstenberg:2008fx}.

\begin{figure}[htbp]
\centering\includegraphics[width=.8\columnwidth]{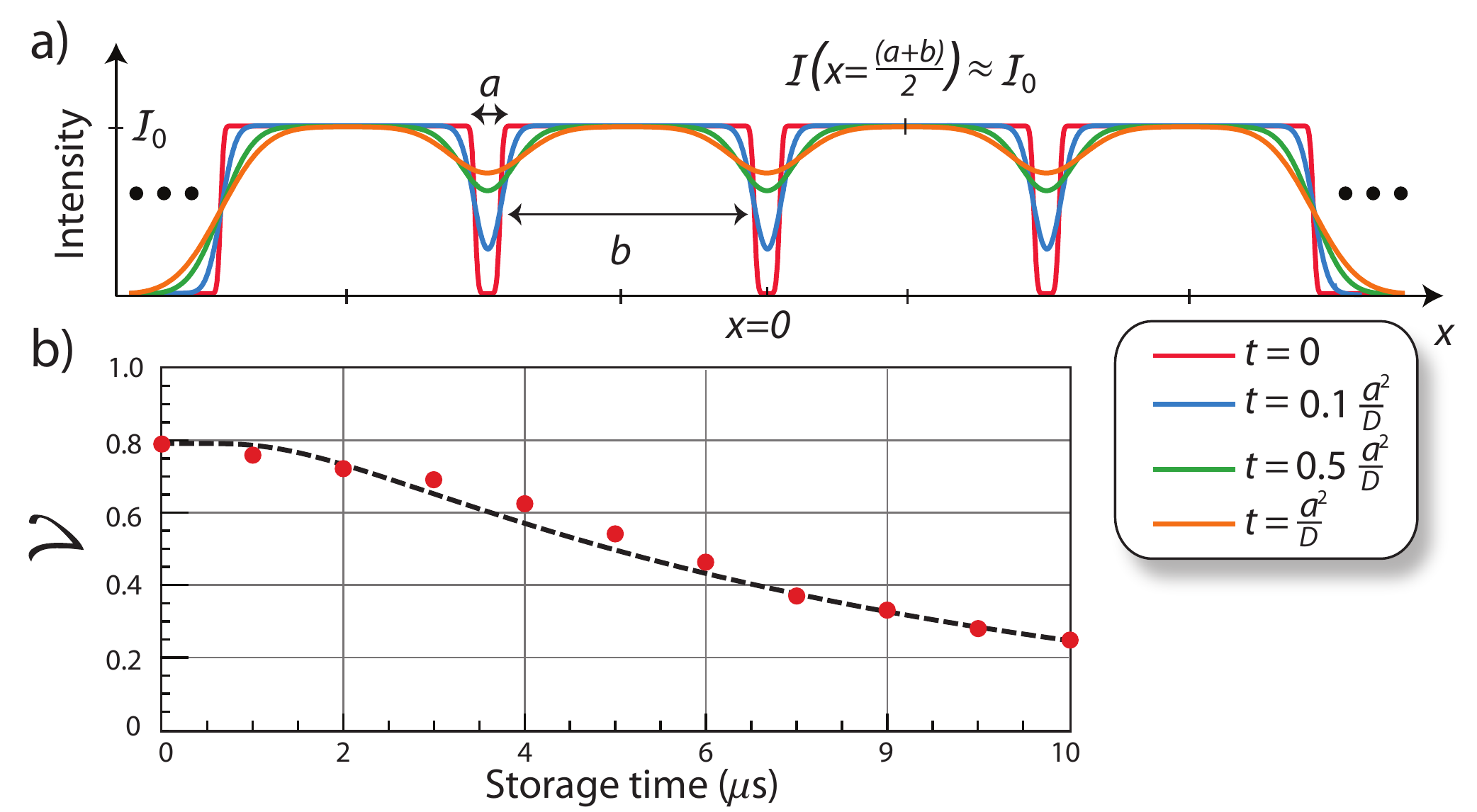}
\caption{\label{fig:contrasttheory} (a) Effect of diffusion on independent channels of width $b$, separation $a$, and diffusion coefficient $D$ at various storage times $t$ assuming no background.  (b) Observed decay of the visibility $\mathcal{V}$ (red points) for a stored resolution target at 1 line pair per millimeter.  The data is compared against the model of Eq.~\ref{eq:contrast2} (black dotted line) assuming a diffusion coefficient $D\approx35$~cm$^2$s$^{-1}$ and a normalized background intensity of 0.1.}
\end{figure}

Taking the diffusive motion of the atoms into account \cite{Firstenberg:2012fu}, we estimate the number of channels $N_{max}$ of a spatially--multiplexed signal that can be stored and independently read--out over time.
For simplicity, we consider the storage of independent channels of width $b$ buffered by channel separations of width $a$ (Fig.~\ref{fig:contrasttheory}).
Defining the origin of our coordinate system at the center of the channel separation, we quantify the expected channel independence under the influence of atomic diffusion according to the fringe visibility, $\mathcal{V}$:
\begin{equation}
\label{eq:contrast}
\mathcal{V}\equiv\frac{\mathcal{I}(x=\frac{a+b}{2},t)-\mathcal{I}(x=0,t)}{\mathcal{I}(x=\frac{a+b}{2},t)+\mathcal{I}(x=0,t).}
\end{equation}
In Eq.~\ref{eq:contrast}, $\mathcal{I}(x,t)$ denotes the recalled intensity at position $x$ after storage time $t$.
In the limit of narrow separations ($\frac{a}{b} \ll 1$) and sufficiently short storage times ($\sqrt{Dt} \ll b$), the retrieved $\mathcal{V}$ can be approximated as
\begin{equation}
\label{eq:contrast2}
\mathcal{V} \approx \bigg(\frac{2}{\mathrm{Erf}(\frac{a}{4\sqrt{Dt}})}-1 \bigg)^{-1}
\end{equation}
where $D$ is the diffusion coefficient for the atomic vapor \cite{Franz:1965vf,Franzen:1959wj}.
$\mathcal{V}$ does not depend on $b$ in this limit because only nearest neighbor channels are expected to meaningfully contribute to the decay of the channel separation for short storage times.

By defining a threshold on the minimum acceptable visibility $\mathcal{V}_{lim}$, we arrive at an analytical solution for the maximum allowable linear channel density $\Lambda$:
\begin{equation}
\label{eq:D}
\Lambda=\bigg(4\sqrt{Dt}~\mathrm{Erf}^{-1}\bigg(\frac{2\mathcal{V}_{lim}}{1+\mathcal{V}_{lim}}\bigg)+b\bigg)^{-1}.
\end{equation}
Given our empirically determined diffusion coefficient $D\approx35$~cm$^2$s$^{-1}$ and a $\mathcal{V}_{lim}$~=~0.9, we estimate that our memory can accommodate a linear channel density of up to $\Lambda\approx$~7~cm$^{-1}$ for up to three readouts assuming a total storage time of 15 $\mu$s (t=15~$\mu$s in Eq.~3).
This density falls well within the resolution afforded by our read beams.
We conclude that, for $\mu$s storage times, the upper bound on $\Lambda$ is limited by atomic diffusion.

\section{Decoherence-induced erasure}
The evolution of the probe optical field $\mathcal{E}$ and atomic ground state coherence $\sigma_{2,3}$ can be described by the coupled Maxwell-Heisenberg equations \cite{Hosseini:2012go}:
\begin{eqnarray}
\label{eq:maxheis}
\partial_t\sigma_{2,3}(z,t) &=& -\bigg(\gamma_{2,3}+i\eta z\bigg)\sigma_{2,3}(z,t)+ig\mathcal{E}(z,t),\nonumber\\
\partial_z\mathcal{E}(z,t) &=& \frac{i\mathcal{N}\Omega}{\Delta_w}\sigma_{2,3}(z,t),
\end{eqnarray}
where $\mathcal{E}(z,t)$ is the slowly-varying electric field of the probe, $\sigma_{2,3}(z,t)$ the $F=\{2,3\}$ ground state spin wave, $\mathcal{N}$ the linear atomic density, g the Raman coupling strength, $\Omega$ the Rabi frequency of the control beam, and $\Delta_w$ the detuning of the control beam from resonance.
If the ground state decoherence rate $\gamma_{2,3}$ is assumed small compared to the time scale of the experiment, the general behavior of the system is to coherently transfer an optical excitation into a ground state spin wave \cite{Hosseini:2012go}.
Accordingly, any decoherence process that damages the spin wave during storage will directly result in loss of the retrieved probe pulse.
\begin{figure}[htbp]
\centering\includegraphics[width=\columnwidth]{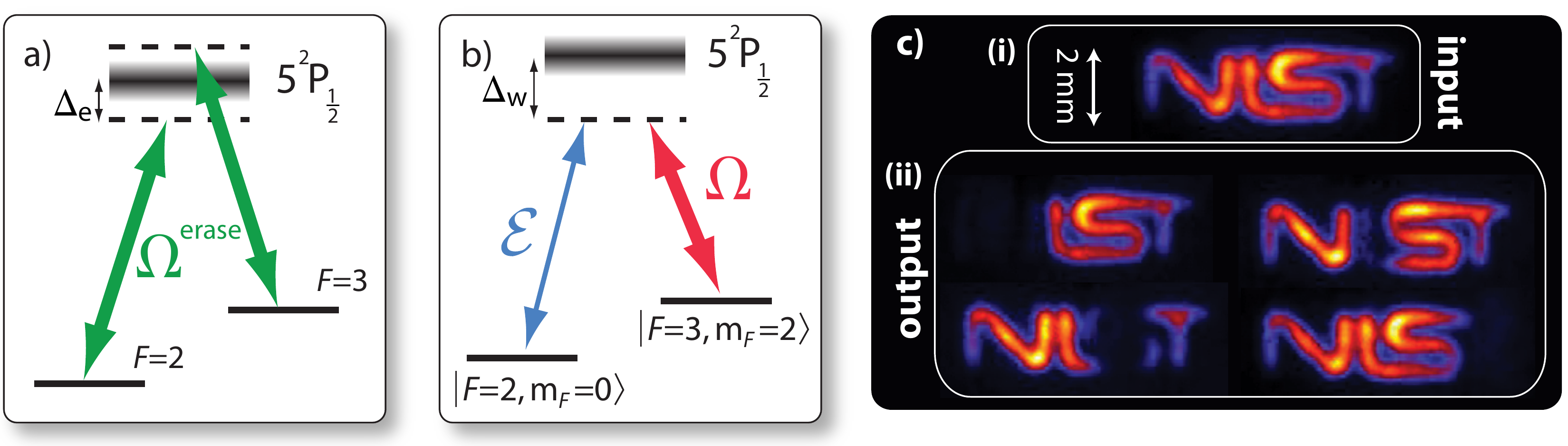}
\caption{\label{fig:fig3} Local erasure of an image.
a) Optical eraser applied between the hyperfine ground state splitting with detuning $\Delta_{e}~\approx$~1.5~GHz.
(b) Raman absorption scheme used to excite the spin wave. Write beam detuning $\Delta_{w}~\approx$~2~GHz.
(c) (i) Retrieved image after 3~$\mu$s of storage without erasure.  (ii) Retrieved images after 3~$\mu$s storage times and local erasure.}
\end{figure}

We intentionally induce decoherence of the ground state spin wave by pulsing a 10~mW ``optical eraser" detuned by $\Delta_e\approx$~1.5~GHz from the D$_1$ resonance during storage (Fig.~\ref{fig:fig3}a).
Choosing equal detunings for both ground states is meant to minimize optical pumping effects in the $\Lambda$-system, allowing us to approximate the storage medium's response as that of a two-level atom.
We therefore expect the characteristic decoherence rate  $\gamma_{2,3}$ to scale with the scattering rate:
\begin{equation}
\mathcal{R}_{sc} \approx \frac{\Gamma}{2}\bigg(\frac{\mathcal{I}}{\mathcal{I}_{sat}}\bigg)\bigg(1+4\bigg(\frac{\Delta_e}{\Gamma}\bigg)^2+\frac{\mathcal{I}}{\mathcal{I}_{sat}}\bigg)^{-1},
\end{equation}
where $\Gamma=2\pi\cdot5.75$~MHz is the $5^2P_{\frac{1}{2}}$ excited state linewidth, $\mathcal{I}_{sat}$ is the resonant saturation intensity of the D$_1$ line, and $\mathcal{I}$ is the optical eraser beam intensity.
The decoherence of the ground state  $\gamma_{2,3}$ therefore depends on the detuning, optical power, and pulse width of the optical eraser.

In all experiments, the write beam is also used as a read beam (Fig.~\ref{fig:fig3}b) in order to attempt the recall of the entire stored image in a given experiment and to ensure identical conditions during storage and retrieval.
We prepare the optical eraser by replacing the optical fiber previously used for the central read beam (``read 2" in Fig.~1b) with an independent diode laser tuned to $\Delta_e\approx$~1.5 GHz.
Fig.~\ref{fig:fig3}c demonstrates how letters of the NIST logo can be individually erased.
All images subject to targeted erasure were captured from individual echo experiments since information stored in erased regions is irreversibly lost regardless of any temporal multiplexing.

\begin{figure}[htbp]
\centering\includegraphics[width=\columnwidth]{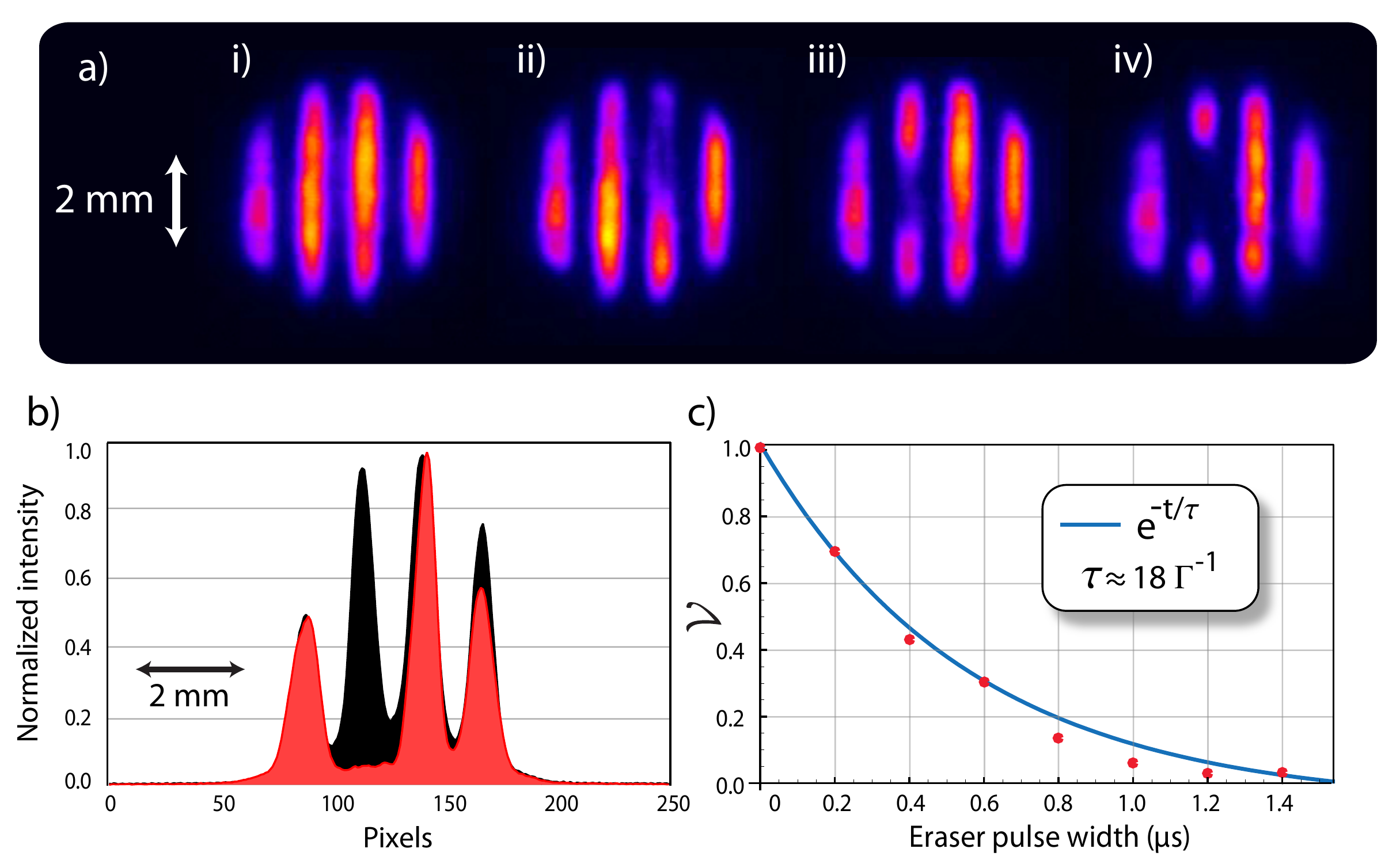}
\caption{\label{fig:fig4}Localized erasure of a stored image.
a) Retrieved fringe patterns (1.5 line pairs per mm) for the case of (i) no optical eraser, (ii) and (iii) a 600~ns eraser pulse, and (iv) a 1~$\mu$s eraser pulse.
b) Profile plot of the retrieved intensity without erasure (black) and after exposure to a 1~$\mu$s eraser pulse (red).
c) Visibility $\mathcal{V}$ of the retrieved fringe patterns (red points) versus pulse width of the optical eraser. The solid blue curve is the modeled decay of $\mathcal{V}$ predicted by the scattering rate of a two-level atom for a detuning of $\Delta$=1.5~GHz.  In all cases the storage time was fixed to 3~$\mu$s.}
\end{figure}

To quantitatively study effects of the eraser, we store a resolution target pattern with a density of 1.5 line pairs per mm (see Fig.~\ref{fig:fig4}a).
Since the edge-coated mirror assembly was mounted on a translation stage, our setup facilitated easily controllable targeting of 2x3 mm regions of interest in the fringe pattern.
In particular, we found that erasure could be strongly localized to a particular region of interest, allowing for the complete removal of a selected fringe with negligible effects on neighboring fringes (Fig.~\ref{fig:fig4}b).
This degree of precision is consistent with the resolution limits achieved in this experiment, suggesting that this erasure technique could be used to delete undesired channels of a densely spatially--multiplexed signal.

To study the time scale necessary for complete erasure, we measured the retrieved local fringe visibility $\mathcal{V}$ subject to a variety of eraser pulse widths.
All fringe patterns were stored for 3 $\mu$s to ensure that any effects due to diffusion were common among all shots.
When evaluating $\mathcal{V}$ in a given shot, the fringe patterns were normalized to the brightest undeleted fringe in order to correct for shot-to-shot laser fluctuations.
Additionally, an average background from the read beam was subtracted from each shot to allow $\mathcal{V}$ to approach zero in the limit of large eraser pulse widths.
The measured decay of $\mathcal{V}$, as well as the expected decay for a two-level atom, are plotted against storage time in Fig.~\ref{fig:fig4}c.  
Using the model of a two-level atom, we estimate a decay rate of $\approx$~18~$\Gamma^{-1}$.
This very simple model shows close agreement with a best-fit exponential decay rate of $\approx$~17~$\Gamma^{-1}$.

\section{Conclusion}
Storage and retrieval of multi-spatial-mode images in an atomic memory relaxes the requirements on storage time for building a scalable quantum network based on quantum repeaters.
We have demonstrated that GEM is suitable for on-demand retrieval of independent optical channels stored in the memory.
To improve the flexibility of the technique, we introduced a method allowing fast erasure of undesired subregions of an image, which in principle could be extended to a spatially (though not temporally) multiplexed quantum signal \cite{Clark:12,Boyer:2008ts}.
We observed that, under our experimental conditions, the spatially addressable resolution for readout and erasure is limited by the imaging system for the read and eraser beams and that the maximum density of information that can be spatially multiplexed is limited by the atomic diffusion of the atoms.
The effects of atomic motion on storage time have recently been shown to be mitigated in GEM experiments performed using cold atomic ensembles released from an elongated magneto-optical trap \cite{Sparkes:2012}.
Given the transverse extent of the trap and the slow expansion of the cloud during time-of-flight, we estimate that similar cold ensembles should be able to support tens of spatial channels on the millisecond time scale, given diffraction-limited read/write beams at 795 nm.
Additionally, we expect that storing the Fourier transform of an image should be robust in the presence of diffusion, similar to results presented in \cite{Vudyasetu:2008wv}.

\section*{Acknowledgments}
This work is supported by the AFSOR and the Physics Frontier Center of the Joint Quantum Institute.

\section*{References}
\bibliographystyle{iopart-num}
\bibliography{biblio}

\providecommand{\newblock}{}
\begin{thebibliography}{10}
\expandafter\ifx\csname url\endcsname\relax
  \def\url#1{{\tt #1}}\fi
\expandafter\ifx\csname urlprefix\endcsname\relax\def\urlprefix{URL }\fi
\providecommand{\eprint}[2][]{\url{#2}}

\bibitem{Braunstein:2005wr}
Braunstein S and Van~Loock P 2005 {\em Reviews of Modern Physics\/} {\bf 77}
  513--577

\bibitem{Kimble:2008uv}
Kimble H~J 2008 {\em Nature\/} {\bf 453} 1023--1030

\bibitem{Sangouard:2011bp}
Sangouard N, Simon C, de~Riedmatten H and Gisin N 2011 {\em Review of Modern
  Physics\/} {\bf 83} 33--80

\bibitem{Lvovsky:2009vg}
Lvovsky A, Sanders B and Tittel W 2009 {\em Nature Photonics\/} {\bf 3}
  706--714

\bibitem{Hosseini:2011iv}
Hosseini M, Campbell G, Sparkes B~M, Lam P~K and Buchler B~C 2011 {\em Nature
  Physics\/} {\bf 7} 794--798

\bibitem{Hosseini:2012go}
Hosseini M, Sparkes B~M, Campbell G~T, Lam P~K and Buchler B~C 2012 {\em
  Journal of Physics B: Atomic, Molecular and Optical Physics\/} {\bf 45}
  124004

\bibitem{Hetet:2008wi}
Hetet G, Longdell J, Sellars M, Lam P and Buchler B 2008 {\em Physical Review
  Letters\/} {\bf 101} 203601

\bibitem{Collins:2007cs}
Collins O, Jenkins S, Kuzmich A and Kennedy T 2007 {\em Physical Review
  Letters\/} {\bf 98} 060502

\bibitem{Lan:2009ts}
Lan S~Y, Radnaev A~G, Collins O~A, Matsukevich D~N, Kennedy T~A~B and Kuzmich A
  2009 {\em Optics Express\/} {\bf 17} 13639--13645

\bibitem{Glorieux:2012uo}
Glorieux Q, Clark J~B, Marino A~M, Zhou Z and Lett P~D 2012 {\em Optics
  Express\/} {\bf 20} 12350--12358

\bibitem{Higginbottom:2012gl}
Higginbottom D, Sparkes B, Rancic M, Pinel O, Hosseini M, Lam P and Buchler B
  2012 {\em Physical Review A\/} {\bf 86} 023801

\bibitem{Lvovsky:2012}
MacRae A, Brannan T, Achal R and Lvovsky A~I 2012 {\em Physical Review
  Letters\/} {\bf 109} 033601

\bibitem{Guo:2013}
Ding D~S, Wu J~H, Zhou Z~Y, Liu Y, Shi B~S, Zou X~B and Guo G~C 2013 {\em
  Physical Review A\/} {\bf 87} 013835

\bibitem{Eisert:2002}
Eisert J, Scheel S and Plenio M~B 2002 {\em Physical Review Letters\/} {\bf 89}
  137903

\bibitem{Furusawa:2010}
Takahashi H, Neergaard-Nielsen J~S, Takeuchi M, Takeoka M, Hayasaka K, Furusawa
  A and Sasaki M 2010 {\em Nature Photon\/} {\bf 4} 178--181

\bibitem{Zhang:2011}
Zhang Y~Q and Xu J~B 2011 {\em Journal of Modern Optics\/} {\bf 58} 593--598

\bibitem{Hosseini:2009fd}
Hosseini M, Sparkes B~M, Hetet G, Longdell J~J, Lam P~K and Buchler B~C 2009
  {\em Nature\/} {\bf 461} 241--245

\bibitem{Vudyasetu:2008wv}
Vudyasetu P~K, Camacho R~M and Howell J~C 2008 {\em Physical Review Letters\/}
  {\bf 100} 123903

\bibitem{Shuker:2008bn}
Shuker M, Firstenberg O, Pugatch R, Ron A and Davidson N 2008 {\em Physical
  Review Letters\/} {\bf 100} 223601

\bibitem{Firstenberg:2008fx}
Firstenberg O, Shuker M, Pugatch R, Fredkin D, Davidson N and Ron A 2008 {\em
  Physical Review A\/} {\bf 77} 043830

\bibitem{Firstenberg:2012fu}
Firstenberg O, Shuker M, Ron A and Davidson N 2012  (\textit{Preprint}
  \eprint{arXiv:1207.6748})

\bibitem{Franz:1965vf}
Franz F 1965 {\em Physical Review\/} {\bf 139} 603

\bibitem{Franzen:1959wj}
Franzen W 1959 {\em Physical Review\/} {\bf 115} 850

\bibitem{Clark:12}
Clark J~B, Zhou Z, Glorieux Q, Marino A~M and Lett P~D 2012 {\em Opt.
  Express\/} {\bf 20} 17050--17058

\bibitem{Boyer:2008ts}
Boyer V, Marino A~M, Pooser R~C and Lett P~D 2008 {\em Science\/} {\bf 321}
  544--547

\bibitem{Sparkes:2012}
Sparkes B~M, Bernu J, Hosseini M, Geng J, Glorieux Q, Altin P~A, Lam P~K,
  Robins N~P and Buchler B~C 2012  (\textit{Preprint} \eprint{arXiv:1211.7171})

\end{thebibliography}
\end{document}